\documentclass[final,5p,times]{elsarticle}

\usepackage{epsfig}
\usepackage{latexsym}
\usepackage[colorlinks=true,linktocpage=true,linkcolor=blue,citecolor=blue,allcolors=blue]{hyperref}
\usepackage{url}
\usepackage[utf8]{inputenc}
\usepackage{enumerate}
\usepackage{color}
\usepackage{xcolor}
\usepackage{amsmath,amssymb}

\usepackage[english]{babel}
\hypersetup{
   colorlinks=true, 
  linktoc=all,     
  linkcolor=blue,  
}

\def\be{\begin{equation}}
\def\ee{\end{equation}}
\def\bearr{\begin{eqnarray}}
\def\eearr{\end{eqnarray}}

\journal{Physics Letters B}

\begin{document}

\begin{frontmatter}

\title{Simultaneous description of high density QCD matter\\ in heavy ion collisions and neutron star observations}

\author[first,second]{Jan Steinheimer}
\affiliation[first]{organization={GSI Helmholtzzentrum f\"ur Schwerionenforschung GmbH},
            addressline={Planckstr. 1}, 
            postcode={D-64291}, 
            city={Darmstadt},
            country={Germany}
}

\affiliation[second]{organization={Frankfurt Institute for Advanced Studies (FIAS), Ruth-Moufang-Str. 1, D-60438 Frankfurt am Main, Germany}}

\author[second]{Manjunath Omana Kuttan}

\author[third,second,fourth]{Tom Reichert}
\affiliation[third]{organization={Institut für Theoretische Physik, Goethe-Universit\"{a}t Frankfurt, Max-von-Laue-Str. 1, D-60438 Frankfurt am Main, Germany}}
\affiliation[fourth]{organization={Helmholtz Research Academy Hesse for FAIR (HFHF), GSI Helmholtzzentrum f\"ur Schwerionenforschung GmbH, Campus Frankfurt, Max-von-Laue-Str. 12, 60438 Frankfurt am Main, Germany}}

\author[fith]{Yasushi Nara}
\affiliation[fifth]{organization={Akita International University, Yuwa, Akita-city 010-1292, Japan}}

\author[third,first,fourth]{Marcus Bleicher}


\begin{abstract}
A combined constraint on the QCD equation of state, at high densities, from connecting neutron star observations to data from heavy ion reactions is presented. We use the Chiral Mean Field Model which can describe neutron star and iso-spin symmetric matter and allows the consistent calculation of the density and momentum dependent potentials of baryons which are then implemented in the UrQMD transport model. In contrast to previous studies, the same equation of state constrained from neutron star properties is also able to describe experimental observables in heavy ion reactions at the HADES experiment. Unlike many other approaches our results are not constraint to densities up to nuclear saturation or perturbative results which allows a continuous description of the equation of state over a large range in baryon density.
\end{abstract}
\end{frontmatter}

Quantum-Chromo-Dynamics (QCD) describes the fundamental interaction governing the physics on sub-nuclear scales. Its fundamental properties are usually studied in high energy collider experiments at RHIC or LHC, e.g. with proton+proton reactions or in the collisions of heavy ions. In reactions at these large collision energies, the net-baryon density is very small and the temperatures are very high. Such a scenario was realized in nature a few microseconds after the Big Bang. This regime of QCD can be probed on the basis of ab-initio calculations or within well controlled perturbative expansions of QCD. Non-perturbative lattice QCD calculations have established a crossover transition from hadronic to partonic matter at vanishing net baryon density \cite{Borsanyi:2010bp,Bazavov:2011nk,Bazavov:2017dus}.

When going to lower collision energies ($\sqrt {s_{NN}}<10$ GeV) the situation becomes dramatically more complicated as one deals with very high net-baryon densities (2-5 times nuclear saturation density $n_0$) and moderate temperatures. This prevents ab-initio lattice QCD calculations due to the well known sign problem and it further prevents many perturbative approaches leaving the study of this regime to effective models of QCD. 

This region of the QCD phase diagram is of great importance and interest. It may contain one of the most exciting features of the phase diagram, namely the critical end point, where the chiral crossover changes into a first-order phase transition. The location of the critical end point (CEP) of QCD, or even its existence, is not yet known. Extrapolations of lattice-QCD results have established that the critical end point may only be located at temperatures and baryon chemical potentials above $\mu_B/T\gtrsim 3$ \cite{Borsanyi:2020fev,HotQCD:2018pds,Vovchenko:2017gkg}. Current estimates within Dyson-Schwinger and FRG approaches suggest $T^{CEP}=80-140$ MeV, $\mu_B^{CEP}=500-800$ MeV \cite{Fischer:2014ata,Fu:2019hdw,Gao:2020qsj,Gunkel:2021oya}. This region of $\mu_B/T\sim5-6$ is similarly favored by Bayesian inference from holographic models \cite{Hippert:2023bel}, Pad\'e type resummations \cite{Basar:2023nkp,Clarke:2024seq}, using finite size scaling of net protons cumulants \cite{Sorensen:2024mry} and lattice QCD extrapolations based on contours of constant entropy density \cite{Shah:2024img}. It is this region that is extremely relevant for the understanding of Neutron Stars (NS) and binary NS mergers and it is in the focus of the experiments at GSI, FAIR, RHIC-BES, HIAF and FRIB.
A direct detection of the critical endpoint is made complicated by the same critical phenomena which define it, namely critical slowing down as well as severe dampening of the critical effects due to finite size and finite lifetime of the systems created in heavy ion reactions \cite{Bluhm:2020mpc,Nahrgang:2011mg}.

In the present paper we suggest a different route to constraining the phase structure of QCD by combining measurements from neutron star mergers and heavy ion collisions with an effective model for the equation of state. 

We will show a very first comparison of simulated results with momentum-dependent potentials, based on an equation of state constrained from neutron star observations, compared to heavy ion data of the HADES experiment. Our results will clearly determine whether a simultaneous description of both regimes of dense QCD matter is possible within one equation of state which is an important step towards a combined and conclusive understanding of dense QCD.

\section{The CMF model and the EoS}

To be able to combine constraints on the EoS from neutron stars with heavy-ion data, a model is required that can provide input for both, based on a limited set of input parameters. For this purpose we will employ the chiral mean field model (CMF) \cite{Papazoglou:1998vr,Steinheimer:2010ib,Motornenko:2019arp}.

A similar version of this model has already been employed to combine information from NS with heavy ion data based on the speed of sound as function of density \cite{Yao:2023yda}.
In the present we will go further and introduce the momentum dependent interaction potentials from the CMF in our microscopic transport simulation. The chiral mean field model is a fully relativistic parity-doublet approach to describe dense QCD matter, including both the baryonic SU(3)-flavor octet and the $\Delta$ baryons with their respective parity partners. The model aims to represent an equation of state (EoS) compatible with empirical data from heavy-ion collisions, astrophysical constraints, and lattice QCD simulations. It effectively integrates scalar and vector mean fields, which impact the baryonic masses $m^*_{b\pm}$ and interactions \cite{Steinheimer:2011ea}. The effective baryon mass is then reduced by the light and strange quark scalar fields $\sigma$, $\zeta$ and reads
\begin{eqnarray}
m^*_{b\pm} = \sqrt{ \left[ (g^{(1)}_{\sigma b} \sigma + g^{(1)}_{\zeta b}  \zeta )^2 + (m_0+n_S m_s)^2 \right]}  \pm g^{(2)}_{\sigma b} \sigma,
\end{eqnarray}
in which $\pm$ stands for positive (negative) parity partners, $g_i^{(j)}$ are the couplings to the scalar fields, $n_S$ is the strangeness of the baryon, $m_0$ represents a bare mass term and $m_s$ is the current quark mass of the strange quark.
The effective chemical potential on the other hand is changed by the vector fields $\omega$ (net baryon density), $\rho$ (net iso-spin density), $\phi$ (net strangeness density) and reads
\begin{equation}
    \mu^*_b=\mu_b-g_{\omega b} \omega-g_{\phi b} \phi-g_{\rho b} \rho,
\end{equation}
where $g_i$ are the couplings to the vector fields. The mean field values are determined by the scalar and vector interactions. Quark degrees of freedom are incorporated in a PNJL-motivated approach having their thermal contribution directly linked to the Polyakov Loop order parameter $\Phi$ and their effective masses also adjusted by the scalar fields. Finally, the model also includes an excluded volume $v_i$ for baryons and mesons, while quarks are assumed to be point-like. A more detailed description of the model and its implementation in UrQMD, including the momentum dependent potentials, can be found in \cite{Steinheimer:2024eha}.

\begin{figure}[t]
  \centering
  \includegraphics[width=0.5\textwidth]{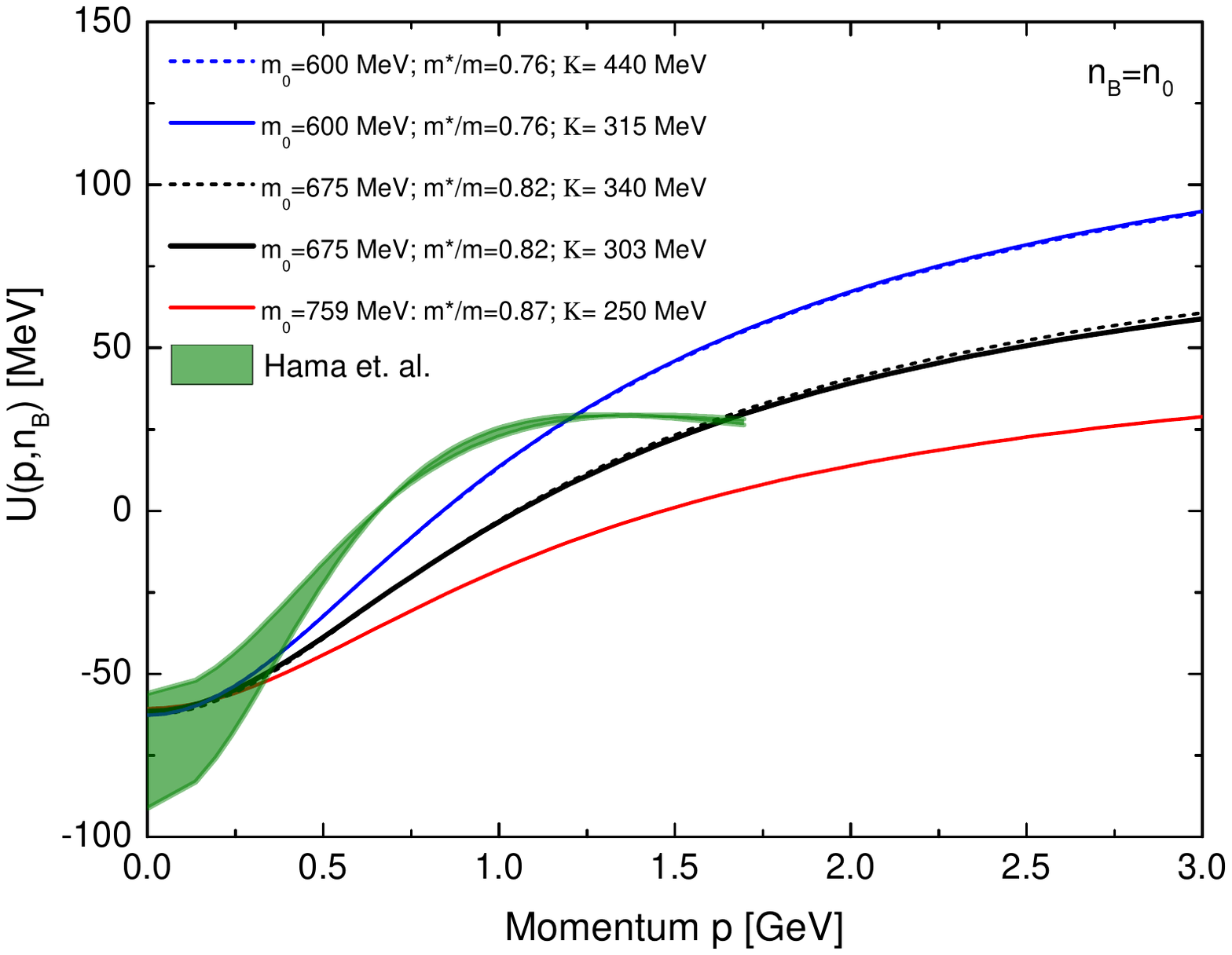}
  \caption{(Color online) Momentum dependence of the single particle potential of nucleons in iso-spin symmetric matter for different parametrizations of the CMF. The green shades area shows momentum dependent Dirac masses extracted from experiment by Hama et.al. \cite{Hama:1990vr}} 
  \label{fig:1}
\end{figure}

One should note that this approach allows us to implement the in-medium potential for all baryons in the CMF model, which includes the lowest baryon octet with its corresponding parity partners as well as the $\Delta$-baryons. 
In principle one could also introduce a potential for the mesons, which requires additional assumptions within the CMF model (see e.g. \cite{Kumar:2025rxj} for recent work within CMF). However, in heavy ion collisions, the pion in-medium effects mainly appear through the excitation and decay of the $\Delta$ which was shown to resembles the dominant contribution to pion nucleon interactions. Any pion interactions beyond the resonance channel (and Coulomb) is usually not taken into account as its effect on the pion properties is negligible \cite{Bertsch:1988xu,Koch:1992zi,Ehehalt:1993px}. 
In our simulations we do include the in-medium modification of the $\Delta$ single particle energy, but not any change in the pion properties in the medium nor do we include any change in the nucleon-$\Delta$ cross section.

Other works within a different transport model have shown that the inclusion of a koan potential can have an effect on the kaon spectra \cite{Song:2020clw}, but the exact treatment will be model dependent. For the current work all mesons are included as free particles.

While in the previous work \cite{Steinheimer:2024eha} the focus was on establishing the UrQMD+CMF model and its implementation, in the following we present a parameter study of the CMF in order to establish whether a simultaneous description of neutron star observables and heavy ion observables is possible, and if so, determine the most suitable parameters.

The parameters that are most relevant for the equation of state as well its momentum dependence are the scalar and vector coupling strengths $g_{\sigma}$ and $g_{\omega}$ as well as the bare mass of the nucleons $m_0$. These are not entirely independent, demanding that the nucleon vacuum masses are reproduced. By fixing the bare mass $m_0$ also the momentum dependence of the single particle energy is fixed. In addition to $g_{\omega}$, the repulsive interaction between hadrons can be modified by their excluded volume parameter $v_i$ and the nuclear incompressibility has been shown to be sensitive to the parameters of the scalar potential. This means we essentially have 4 free parameters in our model but 3 additional constraints, the nuclear saturation density $n_0 = 0.16$ fm$^{-3}$, nuclear binding energy per baryon $E/A-m_N=-16$~MeV and the incompressibility of nuclear matter at saturation density $K$. Although the first two can be considered strict constraints, there is some freedom on the incompressibility. In the following, we will compare five different parametrizations of the CMF model. The resulting momentum dependencies of the single particle potential $U(p,n_B)$, at saturation density, of these 5 scenarios are shown in figure \ref{fig:1}. Here, parameter sets with the same bare mass have the same color. For the blue ($m_0=600$ MeV) and black ($m_0=675$ MeV) curves also a scenario with increased incompressibility (dashed curves) is shown. As expected, a smaller bare mass, resulting in a stronger scalar coupling, leads to a stronger momentum dependence. A larger scalar coupling tends to also require a stronger vector repulsion to be able to reproduce nuclear binding and saturation properties which leads to a higher incompressibility. Figure \ref{fig:1} lists the effective mass of the nucleon at saturation density for all scenarios. The momentum dependence of the single particle energy derived from proton+nucleus scattering experiments is shown for comparison as green band. This band corresponds to two possible scenarios of the momentum dependence of the relativistic Dirac potential extracted in \cite{Hama:1990vr} \footnote{Note, that also in the QMD part of the UrQMD model we will use the relativistic single particle potential from CMF instead of the Schr\"odinger equivalent potential \cite{Weber:1992qc} as the UrQMD model uses the relativistic kinetic energy and the momentum dependent potential can be understood as an effective way of introducing an effective mass in the relativistic kinetic energy.}. For the following comparisons we therefore have 5 different scenarios with 3 different momentum dependencies and a range of incompressibilities. 

\begin{figure}[t]
  \centering
  \includegraphics[width=0.5\textwidth]{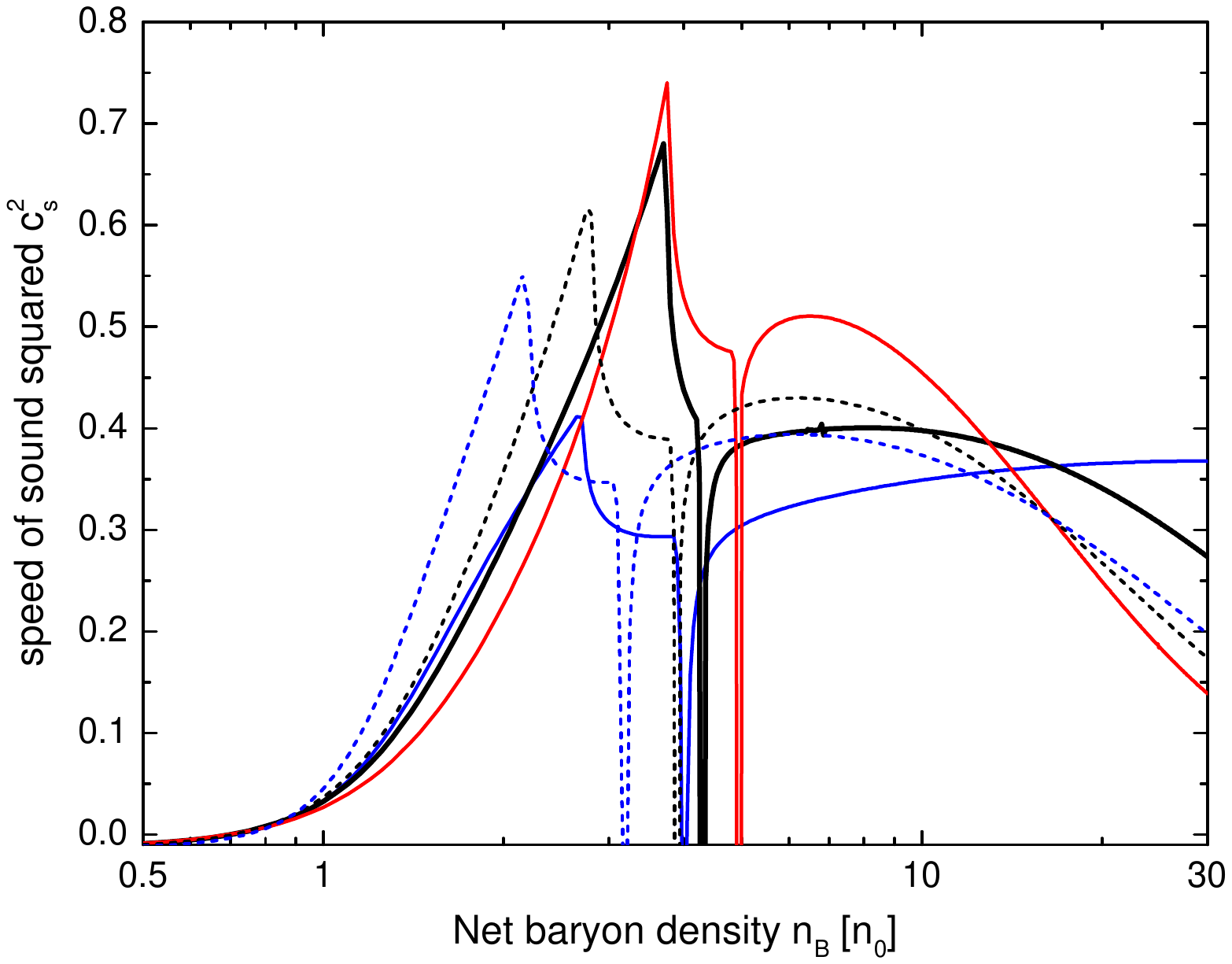}
  \caption{(Color online) Speed of sound squared at $T=0$ in iso-spin symmetric matter for the different parametrizations.} 
  \label{fig:2}
\end{figure}

The equation of state of symmetric nuclear matter, characterized by the speed of sound, for all the different parametrizations of the CMF model is presented in figure \ref{fig:2}. The line colors and styles are the same as in fig. \ref{fig:1}.
 All curves show a characteristic peak of the speed of sound, its position and height depend on the parameters. The strong increase in the speed of sound above nuclear saturation density is caused in part by the vector mean field repulsion but also by the excluded volume of the baryons. The softening of the equation of state is then due to the slow appearance of deconfined quarks which do not interact repulsively. While scenarios with a large incompressibility show a peak at a rather low density, the height of the peak seems to be systematically lower for a strong momentum dependence. All equations of state show a decrease of the speed of sound below the conformal limit of $1/3$ due to the appearance of free quarks that start to dominate at high densities. At asymptotically high densities, the speed of sound will slowly approach the (conformal) limit of a free gas of massless quarks, as expected, by construction.

\section{Neutron star matter}

Using the Tolmann-Oppenheimer-Volkoff (TOV) equation \cite{Oppenheimer:1939ne}, one can relate the equation of state to a unique mass-radius relation of neutron stars. To do so in the CMF model, we perform calculations assuming $\beta-$equilibrium matter, including electrons and muons, and without strangeness conservation.
In the CMF model, the symmetry energy of nuclear matter is controlled mainly by the coupling parameter to the iso-vector field $\rho$: $g_{\rho}$. This coupling is set to $g_{\rho}=4.9$, for all baryons and all scenarios discussed, which leads to a symmetry energy in the range of 30-32 MeV for all cases. This is in the range of the known uncertainties. The remaining density dependence of the symmetry energy is then controlled by the iso-scalar couplings which are discussed in the remainder of the paper. This leads, for example, to a varying value of the slope of the symmetry energy $L$ between 43 and 73 MeV as shown in table \ref{tab:fac}. As the value of $L$ is experimentally not well constrained, this observable does not help us to further pin down the remaining couplings. However, as we have shown, the actual M-R relation is more sensitive on the density dependence of the EoS and thus allows a stronger constraint than obtained from the slope parameter.\\

\begin{table}[h!]
\centering
\begin{tabular}{|c|c|}
\hline
Parameters   & $L$ [MeV] \\
\hline
$m_0=600$ MeV; $m^*/m=0.76$; $K= 440$ MeV & 48\\
\hline
$m_0=600$ MeV; $m^*/m=0.76$; $K= 315$ MeV & 44\\
\hline
$m_0=675$ MeV; $m^*/m=0.82$; $K= 340$ MeV & 55\\
\hline
$m_0=675$ MeV; $m^*/m=0.82$; $K= 303$ MeV & 53\\
\hline
$m_0=759$ MeV; $m^*/m=0.87$; $K= 250$ MeV & 73\\
\hline
\hline
\end{tabular}
\caption{Slope $L$ of the symmetry energy for the different parameterizations used. \label{tab:fac}}
\end{table}

\begin{figure}[t]
  \centering
  \includegraphics[width=0.5\textwidth]{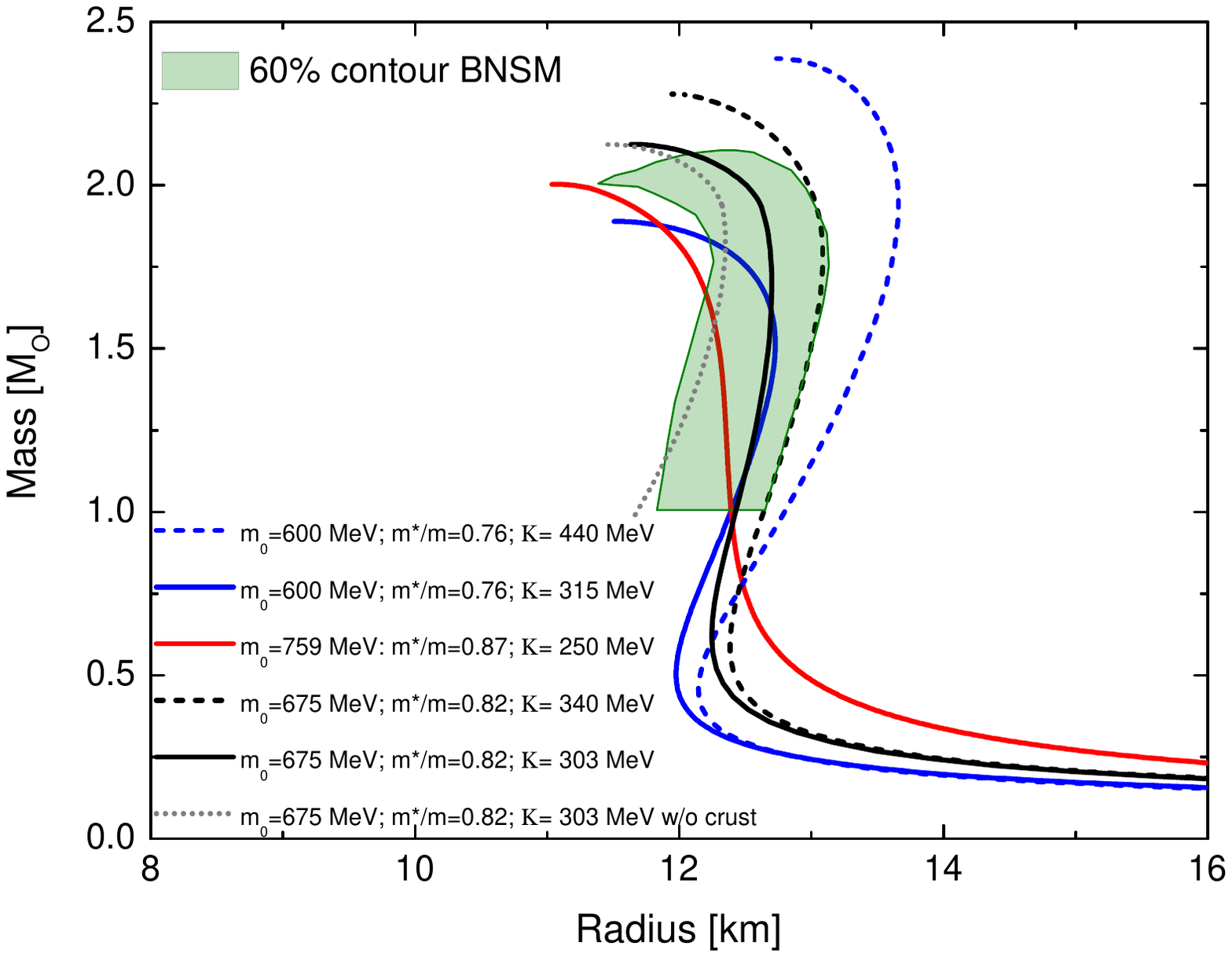}
  \caption{(Color online) Neutron star M-R curves for the different EoS's used. The solid black line gives the best agreement. Too stiff EoS give too large radii and too strong momentum dependencies too small maximum masses.}
  \label{fig:3}
\end{figure}

As the CMF model does not include atomic nuclei as degrees of freedom it will underestimate the pressure of low density matter found in the crust of neutron stars. We therefore need to match the CMF model to a model for the crust at some density below nuclear saturation density. We choose this to be at $\rho_B=0.25 \rho_0$. At this density the pressure in both models becomes similar and allows for a good matching. This combined EoS is then used to solve the TOV equation.
Using the well established DD2 \cite{Lalazissis:2005de} equation of state for the crust we obtain five different mass radius curves for the different scenarios. These are shown as lines in figure \ref{fig:3}. The colors and styles of the lines are the same as in the previous figures. The green shaded area is a constraint from a binary neutron star merger \cite{Altiparmak:2022bke} which is also consistent with NICER constraints \cite{Miller:2019cac,Riley:2019yda,Raaijmakers:2019qny,Miller:2021qha,Riley:2021pdl,Raaijmakers:2021uju}. As one can see the solid black line gives the best description of the constraint which corresponds to an intermediate momentum dependence and a nuclear incompressibility of $K = 303$ MeV. This value for the nuclear incompressibilities is larger than early constraints from the giant dipole resonance ($K = 210 \pm30$ MeV) found in  \cite{Blaizot:1980tw}. However, a more recent and widely accepted study, that takes into account surface effects on the giant dipole resonance \cite{Stone:2014wza}, obtained a different range ($250 < K < 315$ MeV) which is consistent with our value of $K \approx 303$ MeV.

In the following we will refer to results with this parametrization as the \textit{best fitting parametrization} and always show it as a black solid line. 
Using the best fitting parametrization, where ${m_0=675}$~MeV and $K = 303$~MeV, we also calculated the symmetry energy $E_{sym}=31$ MeV and slope of the symmetry energy $L=53$ MeV, which are both well within the range of experimental observations \cite{Li:2021thg}.

To study the sensitivity of our M-R-results on the matching condition, we have added another solution of the TOV equation to figure 3 as grey dotted line. Here, we have solved the TOV equation using only the CMF model without any matching to the crust as an extreme, physically unrealistic, case. As we can see, the maximum mass and radius is only marginally influenced by the crust and the influence of the crust becomes relevant only for neutron stars with masses below 1.5 solar masses. In general, the radius then will decrease without a crust. The general conclusions of our study do not change even without any crust and the best fitting parametrization still describes the M-R constraints well.

\section{Results from HIC}

In heavy ion collisions the equation of state is mostly inferred through the directed $v_1$ and elliptic flow $v_2$, being the first and second order Fourier coefficient in the expansion of the azimuthal angular distribution \cite{Voloshin:1994mz}, of protons \cite{Steinheimer:2022gqb,Sorensen:2023zkk}. In the energy regime probed with the SIS18 accelerator the generation of elliptic flow follows an intricate interplay of interactions via the potential and collisions \cite{Reichert:2024ayg}. The HADES experiment at GSI has recently measured highly differential data of proton and light nuclei elliptic flow \cite{HADES:2020lob,HADES:2022osk}. 
To relate the different CMF parametrizations to the heavy ion data by HADES, we will employ the UrQMD model which has recently been extended to incorporate a momentum dependent potential from CMF \cite{Steinheimer:2024eha}. Since the procedure has not been changed in the current work we refer to \cite{Steinheimer:2024eha} for the description of the model.

In principle one has to reduce the elastic cross sections for any reactions for which also the potential interaction is included at small relative momenta. Most studies with in-medium elastic cross sections where done at much lower beam energies, where a change in the cross section at a few hundred MeV will have a significant impact on the systems dynamics. The degree of the modification however, is model dependent and still under discussion (see e.g. \cite{Li:1993rwa,Li:1993ef,Sammarruca:2005tk,Zhang:2007zzs,Zhang:2010jf,Mao:1994zza}).
A recent study found that already at a center of mass kinetic energy of 450 MeV, the in-medium reduction factor is above 0.9 \cite{Coupland:2011px}. This means that for spectator-spectator and participant-spectator scatterings at the energies we consider the in-medium reaction can be ignored and even for participant-participant scattering the effect should be small.
Previous results from UrQMD simulations have shown that an effective (global) in-medium reduction factor for heavy ion collisions at beam energies above $E_{\mathrm{lab}}> 1 A$ GeV quickly approaches 1 \cite{Li:2022wvu}. Also in \cite{Li:2018wpv}, for momenta above $p\approx 600$ MeV, the reduction factor becomes unity. This means, that the in-medium cross sections are only relevant for collisions at beam energies of a few hundred MeV but become irrelevant for the higher beam energies discussed in the current manuscript. In our simulations, we do not include any in-medium modification of the elastic NN cross section.

\begin{figure}[t]
  \centering
  \includegraphics[width=0.5\textwidth]{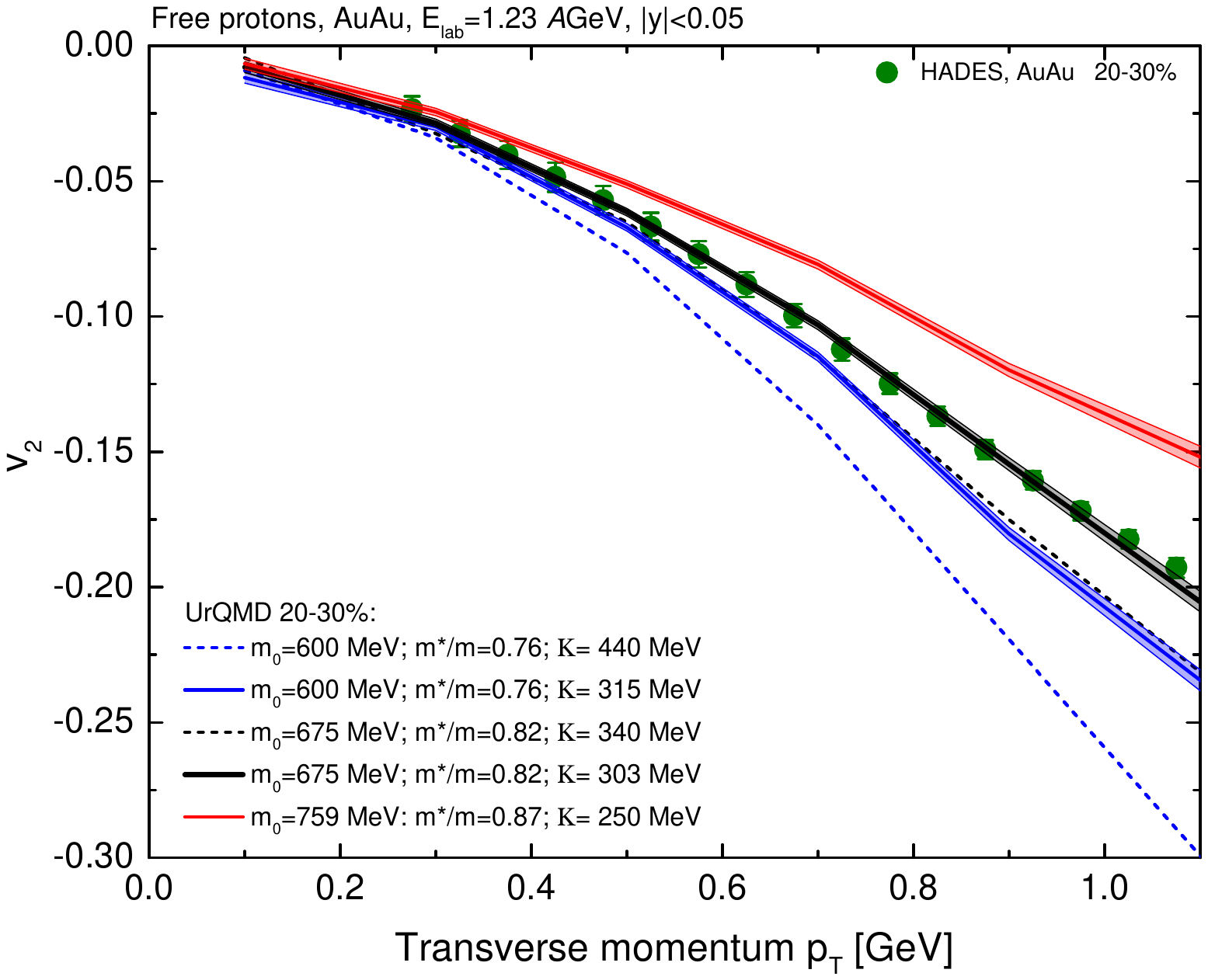}
    \includegraphics[width=0.5\textwidth]{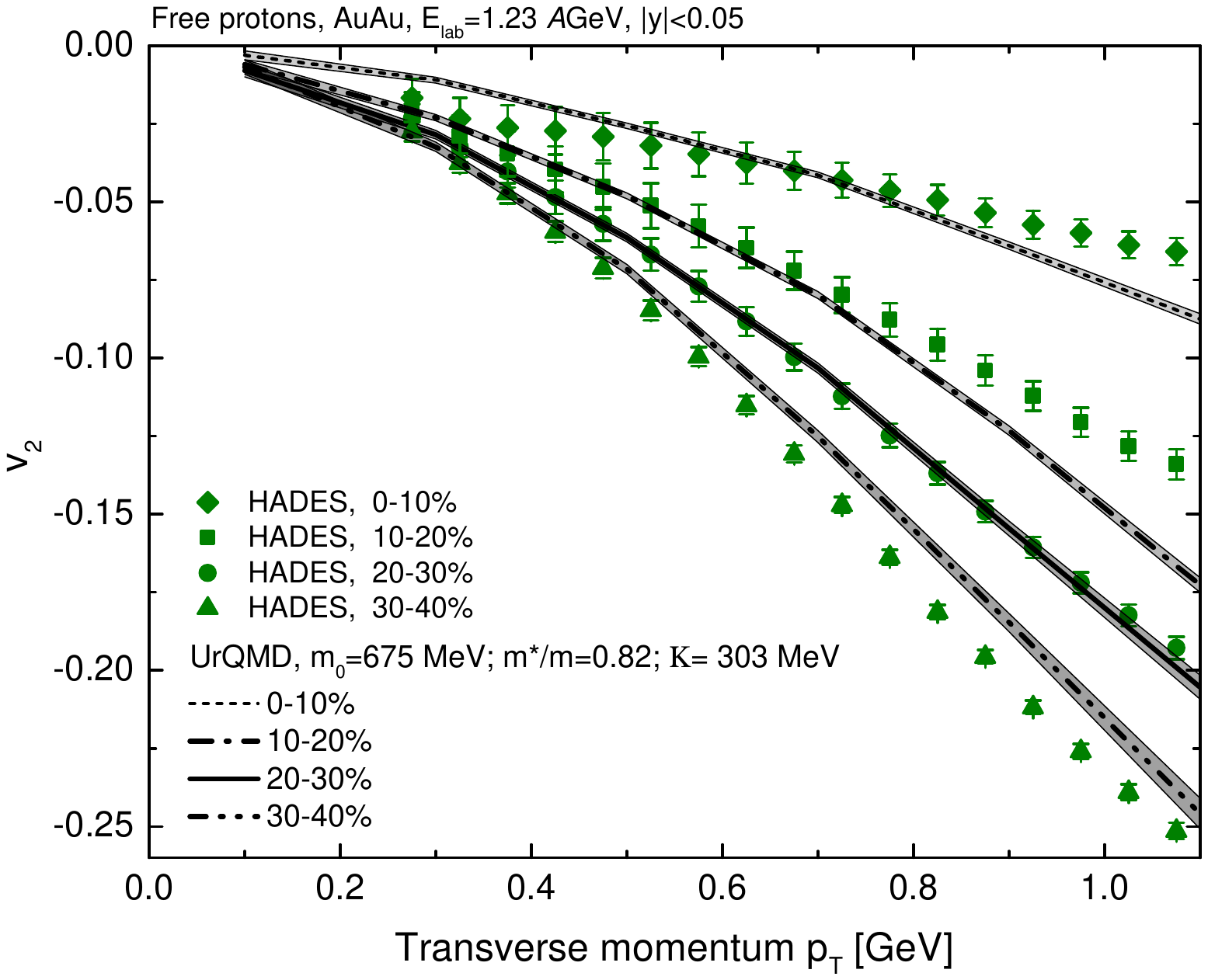}
  \caption{(Color online) Proton $v_2$ as function of transverse momentum for the different EoS compared to HADES data (upper panel). Proton $v_2$ as function of transverse momentum for the best fitting EoS (black line) for different centralities compared to HADES data (lower panel).}
  \label{fig:4}
\end{figure}

\begin{figure}[t]
  \centering
  \includegraphics[width=0.5\textwidth]{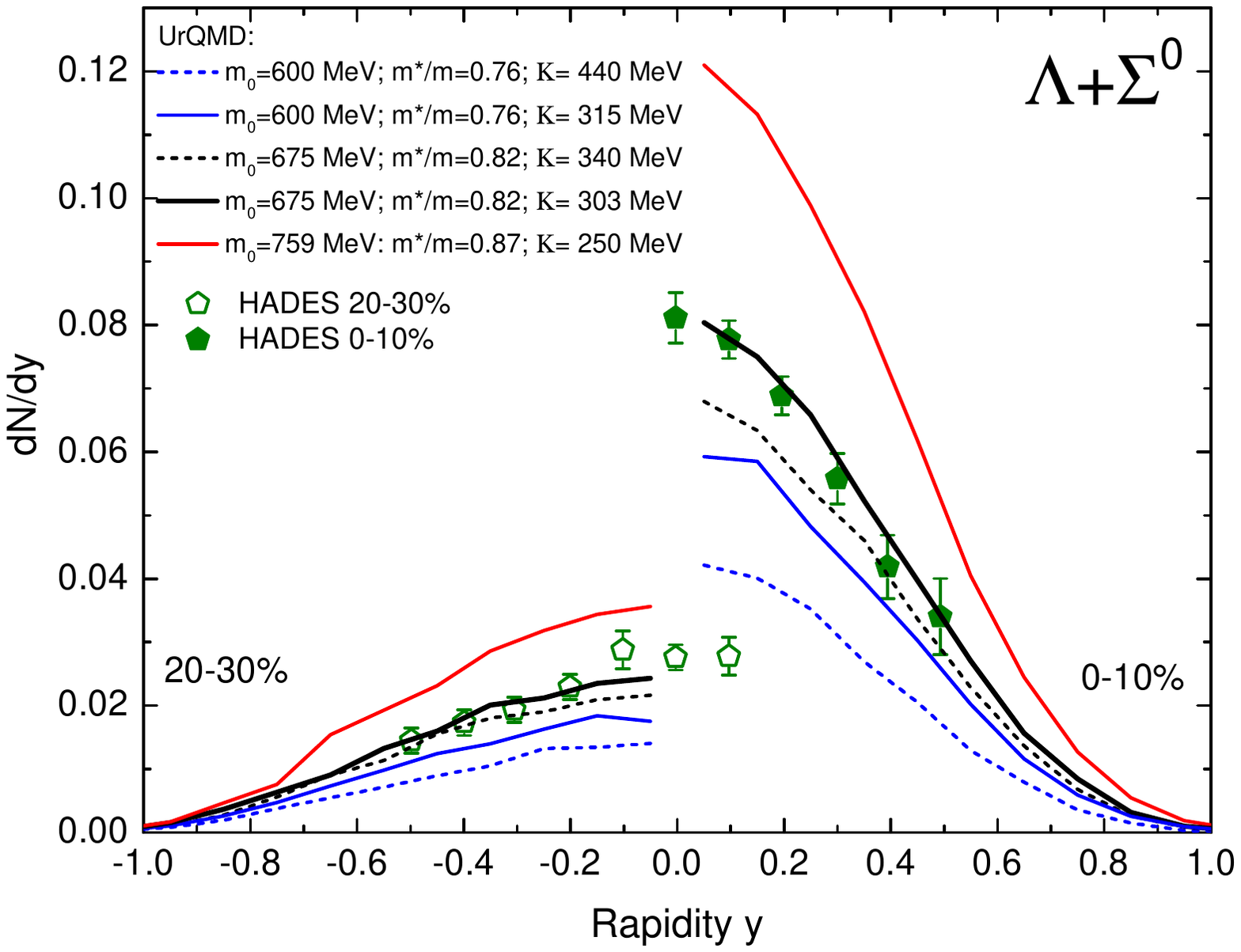}
  \includegraphics[width=0.5\textwidth]{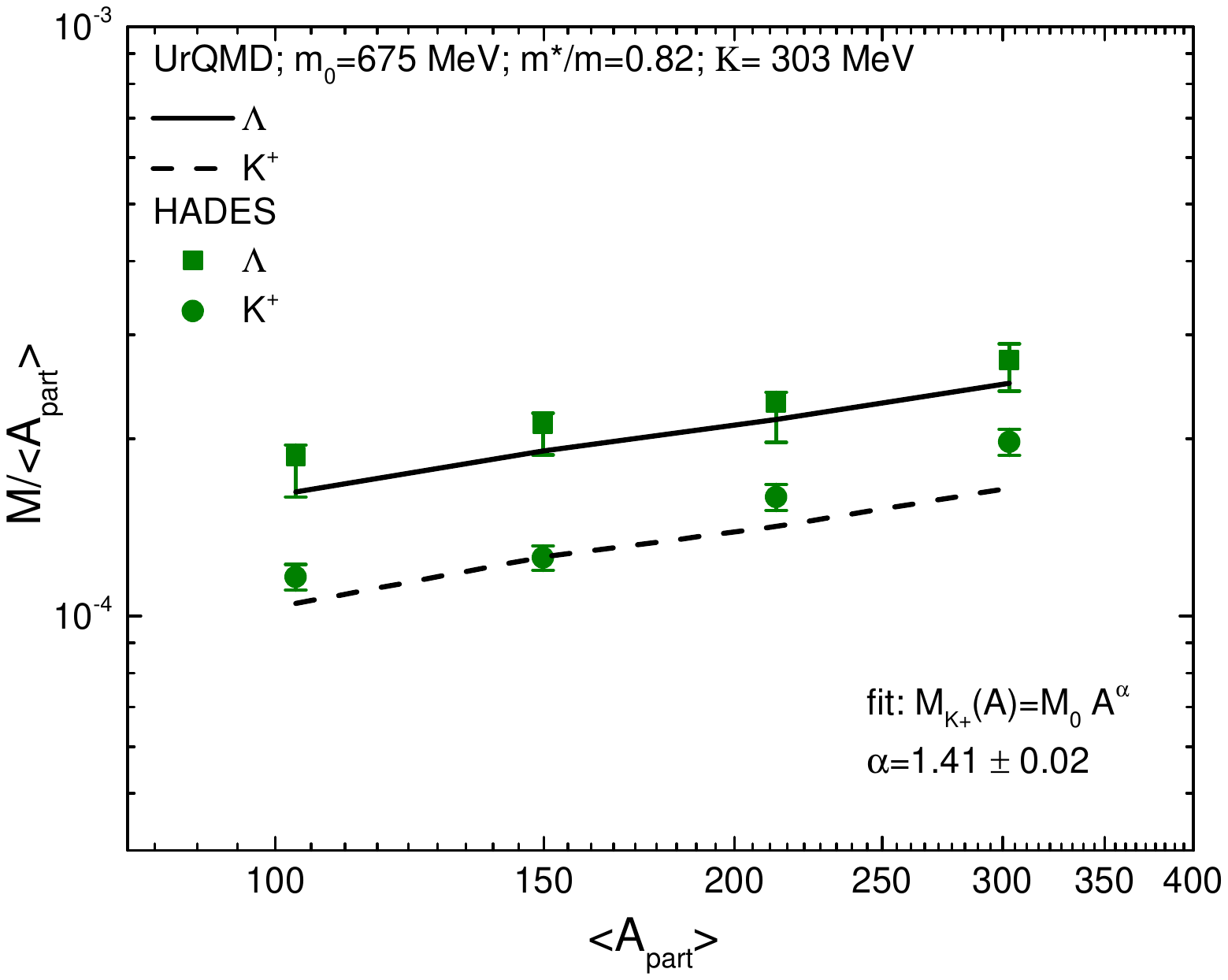}  
  \caption{(Color online) The centrality dependence of $\Lambda$ and $K^+$ production compared to HADES data (upper panel).  $\Lambda$ and $K^+$ rapidity distributions compared to HADES data (lower panel).}
  \label{fig:5}
\end{figure}

In the upper panel of figure \ref{fig:4} we compare results of proton elliptic flow calculated with the different parametrizations of the CMF as a function of transverse momentum $p_T$ in 20-30\% peripheral Au+Au collisions \footnote{For the simulations we used the, Glauber model based, impact parameter ranges given by the HADES collaboration in \cite{HADES:2017def}. These have been shown to correspond to similar centrality classes as obtained from UrQMD simulations \cite{OmanaKuttan:2023cno}.
This method of determining the centrality is not exactly the same as the experimental method. However, it is as close to the experimental setup as one can reasonably get, without having to recreate the whole experimental procedure including a detector simulation.} with the recent HADES measurements. The colors and styles of the lines are the same as in the previous figures. Very clearly the best fitting parametrization obtained from neutron star mass-radius constraints also yields the best description of proton elliptic flow data in 20-30\% peripheral AuAu collisions at $1.23A$ GeV kinetic beam energy. In comparison to the other parametrizations one can conclude that a larger incompressibility turns $v_2$ more negative, as well as a stronger momentum dependence.

While the best fitting parametrization very well describes the elliptic flow in 20-30$\%$ central AuAu collisions at the HADES experiment, it is also important to provide a comparison to the other centrality bins available from HADES. 
We therefore use the best fitting parametrization to evaluate the elliptic flow as a function of transverse momentum for 4 different centrality classes shown in the lower panel of figure \ref{fig:4}. The results show a generally good agreement between the UrQMD model calculations using the best fitting parametrization of the CMF with measured data on elliptic flow.
However, the quality of the agreement varies with centrality. For the most central collisions a deviation for small transverse momenta is observed which is still within the range of the experimental uncertainty.
While the deviation from the experimental data can have different reasons, one should note that our centrality selection and reaction plane reconstruction in the model is not exactly the same as in the experiment which may cause part of the deviation.  

In addition to flow observables, the production of strange hadrons, specifically Kaons, was shown to be sensitive to the equation of state. Although the total yield of Kaons and hyperons depends on both the equation of state and the momentum dependence of the corresponding potentials, the centrality dependence of Kaon production was shown to be sensitive only to the equation of state \cite{Hartnack:2011cn,Hartnack:2005tr,Hartnack:1993bq,Hartnack:1993bp}. The upper part of figure \ref{fig:5} shows the rapidity distributions of $\Lambda$'s for two different centrality bins, using the five CMF parametrizations discussed above. The hyperon rapidity distribution shows a very strong sensitivity to the momentum dependence as well as the equation of state. One clearly observes again, that the best-fit parametrization shown as black line gives the best description of the HADES data \cite{HADES:2018noy} for both centrality bins. 

The total yield of strange hadrons, $\Lambda$ and $K^+$, per number of participants $A_{part}$ as function of the number of participants is presented in the lower panel of figure \ref{fig:5}. As one can see the UrQMD model with the best fitting parametrization gives a good description of the centrality dependence of both hadron yields. The $K^+$ yield is slightly underestimated while the hyperon yield is better described. This is an unexpected result as hyperons and Kaons are produced together and strangeness conservation would imply that either both yields are well described or neither. The discrepancy may be due to the fact that UrQMD also underestimates the ratio of $K^+/K^0$ which is a consequence of the violation of isospin symmetry in strangeness production in elementary SU(3)$_f$ QCD processes \cite{Reichert:2025znn}. 
Nevertheless, one can fit the centrality dependencies, i.e. the number of Kaons and hyperons as function of number of participant simultaneously with a simple function:

\begin{equation}
    M_{h}(A_{part})=M_0 A^{\alpha}_{part}
\end{equation}

Where $M_{h}(A)$ is the number of Kaons or $\Lambda+\Sigma^0$, $A_{part}$ is the number of participants and $\alpha$ is a parameter which quantifies the centrality dependence.
The resulting parameter $\alpha=1.41$ is in good agreement with the HADES result and with previous results from the KaoS and FOPI collaborations \cite{Forster:2007qk,FOPI:2007usx}.

\begin{figure}[t]
  \centering
  \includegraphics[width=0.5\textwidth]{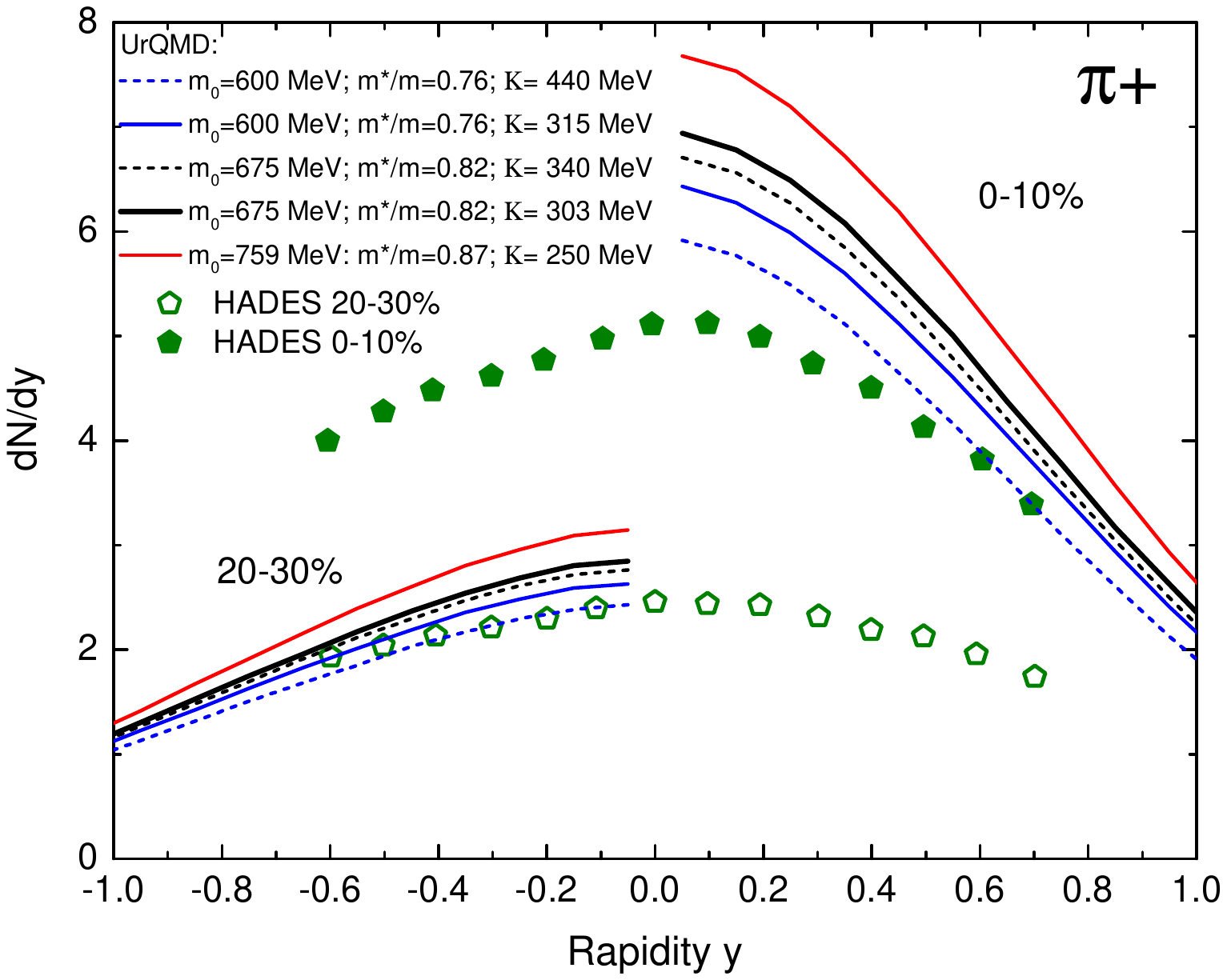}
    \includegraphics[width=0.5\textwidth]{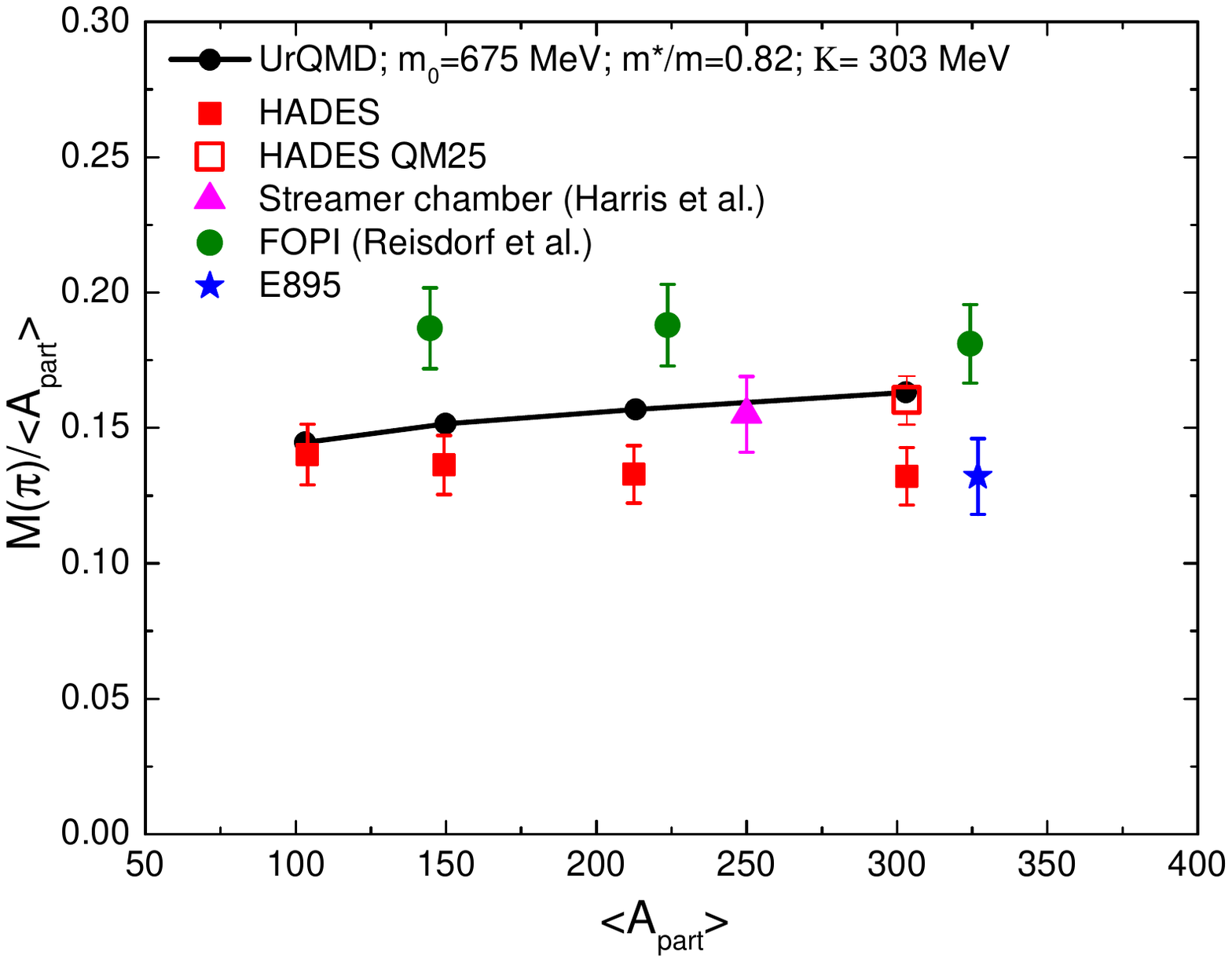}
  \caption{(Color online) Pion rapidity distributions compared to HADES data (upper panel). The centrality dependence is consistent with world data (lower panel).}
  \label{fig:6}
\end{figure}

In addition to the Kaon number, the pion number was also shown to be a sensitive probe of the momentum dependence of the potential \cite{Aichelin:1987ti,Hong:2013yva}. Recent HADES data however have been overestimated by essentially all baseline transport model simulations \cite{HADES:2020ver}.
The deviations of many model calculations from the measured pion multiplicity at the HADES experiment is an interesting aspect which needs to be understood. In its current form the UrQMD model (without momentum dependent potentials) gives similar results on the pion production and rescattering processes as most other transport models shown in \cite{PhysRevC.100.044617,PhysRevC.109.044609,TMEP:2022xjg}.

A modification of the inelastic nucleon-$\Delta$ cross section was discussed as possible solution of this problem \cite{Godbey:2021tbt,Kummer:2023hvl}. Here, it was shown that the pion yield can be reduced by a significant change in the in-medium nucleon-$\Delta$ cross section. 

Figure \ref{fig:6} (upper panel) shows a comparison of the positively charged pion dN/dy from or simulations with the published HADES data in two different centrality bins. A deviation from the HADES data is clearly visible for all parametrizations.

 A more balanced picture emerges as one compares the centrality dependence of the pion production of different experiments with the best fitting parametrization (black) in the lower panel of figure \ref{fig:6}.
 In this figure we also included the newly corrected HADES results that where shown at the Quark Matter 25 conference as \cite{posterqm25} red open square. In this re-analysis of the HADES AuAu data, it was found that an improved efficiency correction leads to an increase of the observed pions which is now in very good agreement with our simulations as can be seen in figure 6. 
 Again, one observes that the best-fit is compatible with the available data within their errors. The resolution of the deviation from previous model calculations to the HADES data may be therefore understood as a combination of including momentum dependent potentials and better efficiency corrections by the experiment. We can therefore say, that our simulations without any additional pion in-medium modifications do very well describe the HADES data on pion production and no in-medium modifications of the pion are necessary.

\begin{figure}[t]
  \centering
  \includegraphics[width=0.5\textwidth]{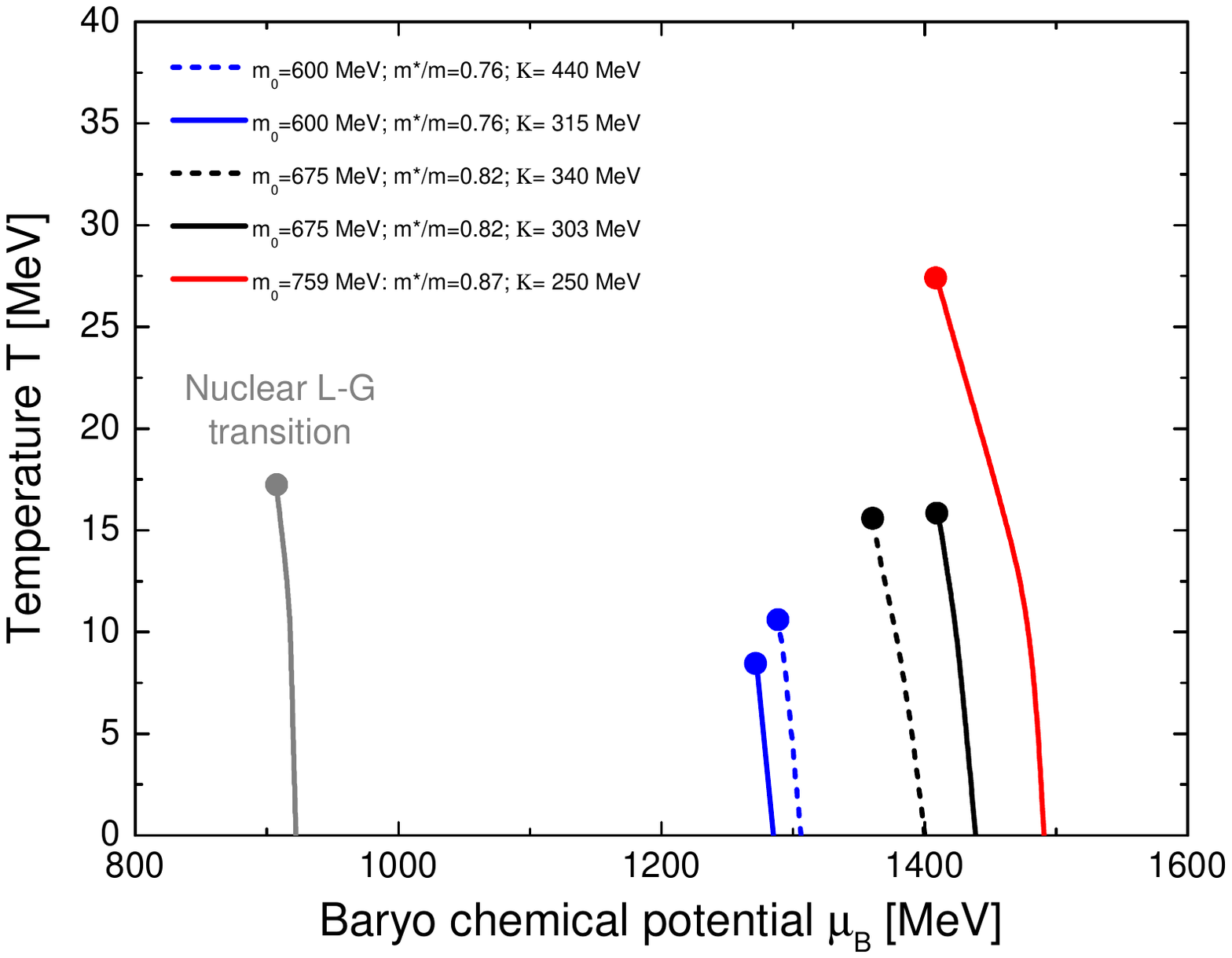}
  \caption{(Color online) Different phase diagrams for the various parametrizations. A larger momentum dependence tends to decrease the critical chemical potential and temperature.} 
  \label{fig:7}
\end{figure}

\section{Discussion}

It was shown how a consistent description of the equation of state (and momentum dependence of the nuclear potential) can be achieved for the description of neutron star properties as well as heavy-ion reactions. To do so we used the Chiral Mean Field Model to calculate several parametrizations of the iso-spin symmetric as well as neutron star equation of state. The parameters of the model where fixed by new constraints on the mass-radius relation from neutron stars. The resulting equation of state is able to describe several experimentally measured observables sensitive to the equation of state and its momentum dependence. A small discrepancy in the pion production at the HADES experiment is still visible. 

The resulting parameters of the best fit can be used to draw a corresponding phase diagram of dense matter which is shown in figure \ref{fig:7}. The well known nuclear liquid-gas transition is shown as grey line. The different colored lines correspond to the chiral transitions in the CMF for the different parameter sets used in our study. The black solid line corresponds to the best fitting curve. Due to the large repulsion at high densities in the CMF model, the critical endpoint for all the scenarios is rather low $T_{CEP}<30$~MeV. The values for the best fitting parametrization are : $T_{CEP}= 16$ MeV and $\mu_{CEP}=1410$ MeV. These results for the location of the QCD critical endpoint are significantly different from e.g. recent predictions of functional methods \cite{Fu:2019hdw,Gao:2020qsj}. The reason for the low critical temperature in the CMF model is the strong short-range baryon repulsion at high density, which is a necessary ingredient to describe the compactness of observed neutron stars. It would be therefore interesting to understand whether the CMF model is missing an important aspect of the QCD transition which drives the transition in the FRG calculations or whether the FRG calculations are missing important aspects of many-baryon interactions.
A more in-depth comparison of the CMF phase structure and thermodynamics to the FRG results is an interesting future project which may help understand the differences.

The resulting CEP is at such a low temperature that it will not be reached by any planned heavy ion experiments and unlikely by neutron star mergers, as shown in \cite{Most:2022wgo}.

The current analysis presents a way forward in how results from astrophysical observations like neutron star masses and binary neutron star mergers can be combined with heavy-ion observables to constrain the high density and temperature QCD equation of state. It also shows a path forward on how such analyses can and should be improved in several ways in the future. This includes the implementation of an explicit dependence of the equation of state on the scalar density, and thus the temperature, which would be naturally included in a relativistic description of the MD-part with a scalar and vector density. 
In addition, another independent and direct constraint for the symmetry energy, and therefore EoS parameters sensitive to the isospin density, from heavy ion reactions would be an important next step. Within the CMF framework that we have laid out, such an extension is straight forward and will be done in a future work. 
Finally, a more complete set of data, including results from other experiments can be used in a statistical inference that can constrain the CMF parameters in a more quantitative way that also includes uncertainties.

\section*{Acknowledgements}
 J.S. and T.R. thank the Lawrence Berkeley Lab for its hospitality and Volker Koch for helpful discussions.
 T.R. acknowledges support through the Main-Campus-Doctus fellowship provided by the Stiftung Polytechnische Gesellschaft Frankfurt am Main (SPTG). T.R. thanks the Samson AG for their support. M.O.K. was supported by the ErUM-Data funding line from BMBF through the KISS project. The computational resources for this project were provided by the Center for Scientific Computing of the GU Frankfurt and the Goethe-HLR.

\bibliographystyle{elsarticle-num} 
\bibliography{main}

\end{document}